# All-Optical tunability of metalenses infiltrated with liquid crystals

Giovanna Palermo,[†,#] Andrew Lininger,[‡,#] Alexa Guglielmelli,[†] Loredana Ricciardi,[ʄ] Giuseppe Nicoletta,[†] Antonio De Luca,[§] Joon-Suh Park,[‖,⊥] Soon Wei Daniel Lim,[‖] Maryna L. Meretska,[‖] Federico Capasso,[*,‖] and Giuseppe Strangi[*,‡,†]

[†]*Department of Physics, NLHT-Lab, University of Calabria and CNR-NANOTEC Istituto di Nanotecnologia, 87036-Rende, Italy*

[‡]*Department of Physics, Case Western Reserve University, 2076 Adelbert Rd, Cleveland, Ohio 44106, USA*

[ʄ]*CNR-NANOTEC Istituto di Nanotecnologia, 87036-Rende, Italy*

[§]*Department of Physics, University of Calabria and CNR-NANOTEC Istituto di Nanotecnologia, 87036-Rende, Italy*

[‖]*Harvard John A. Paulson School of Engineering and Applied Sciences, Harvard University, Cambridge, MA 02138, USA*

[⊥]*Nanophotonics Research Centre, Korea Institute of Science and Technology, Seoul 02792, Republic of Korea*

[#]*Denotes equal contribution to this work*

*E-mail: capasso@seas.harvard.edu

*Email giuseppe.strangi@case.edu

## Abstract

Metasurfaces have been extensively engineered to produce a wide range of optical phenomena, allowing exceptional control over the propagation of light. However, they are generally designed as single-purpose devices without a modifiable post-fabrication optical response, which can be a limitation to real-world applications. In this work, we report a nanostructured planar fused silica metalens permeated with a nematic liquid crystal (NLC) and gold nanoparticle solution. The physical properties of embedded NLCs can be manipulated with the application of external stimuli, enabling reconfig- urable optical metasurfaces. We report all-optical, dynamic control of the metalens optical response resulting from thermo-plasmonic induced changes of the NLC solution associated with the nematic-isotropic phase transition. A continuous and reversible tuning of the metalens focal length is experimentally demonstrated, with a variation of 80 $\mu$m (0.16% of the 5 cm nominal focal length) along the optical axis. This is achieved without direct mechanical or electrical manipulation of the device. The reconfigurable properties are compared with corroborating numerical simulations of the focal length shift and exhibit close correspondence.

# Introduction

Traditionally, the propagation of light is controlled by refractive optical elements.[1] Recently, a paradigm shift has been realized through the introduction of optical metasurfaces, enabling the development of flat optical components by controlling the propagation of light at the nanoscale.[2–10]

Once a given frequency of interest has been set, the desired optical response can be engineered by manipulating the geometry and position of sub-wavelength structures (meta-elements) on the surface. Each meta-element of the metasurface acts as an individual resonator or truncated waveguide, resulting in local control over the transmitted phase and amplitude at high spatial resolution.[11] These advances have led to the fabrication of ultra-thin metalenses with focusing properties matching or exceeding those of typical refractive lenses.[12–14] Metasurfaces have been implemented in a range of passive (non-tunable/non-dynamic) applications, such as focusing and imaging,[15,16] beam conversion,[17] and holography.[18] However, these devices are typically created to support a fixed electromagnetic response which is determined at the time of fabrication. This limitation can be overcome by designing reconfigurable systems, or metasurface devices supporting a dynamically modifiable optical response. Post-fabrication tunability can unlock a range of potential applications, including varifocal imaging, beam steering, dynamic aberration correction and dynamic holography. [19,20] The dynamic response is typically controlled by an external stimulus; common control methods for creating reconfigurable systems involve thermal, electrical, chemical, or mechanical action.[21–24]

Nematic liquid crystals (NLCs) are well known materials in which the refractive index can be modified through the application of an external stimulus, such as an electric or magnetic field.[25] In a typical NLC cell, the liquid crystal is infiltrated between two glass slides that have been properly functionalized in order to obtain the desired molecular alignment.[26] NLCs are known to exhibit a tunable refractive index due to reorientation of the molecular alignment. This tunable refractive index can be utilized to achieve active reconfigurable control over optical metasurface devices. One recent technique to integrate NLCs and metasurfaces is to replace one of the two glass plates in a conventional NLC cell with a metasurface device. The far field optical properties of the metasurface, including diffractive lensing, can then be modified by electrically driving the liquid crystal orientation state, similar to the operation principle of an LC display.[27–29] Additionally, NLC-encapsulated metasurface devices have been programmed as individually addressable pixels. With this technology, electrical mod- ification of adjacent cells, resulting in a manipulation of the local phase, has been used to dynamically create arbitrary holographic patterns.[30] Proceeding beyond these approaches, which utilize a thick LC bulk, in this work we present a metalens device which achieves dynamic and reversible reconfiguration by infiltrating NLC molecules into the space between the nanopillars composing the metasurface. In this scheme, the metasurface creates an initial phase profile for light focusing and the infiltrated LC provides tunability. Having a thick LC superstrate, as in previously published devices, potentially introduces a large transmit- ted phase accumulation in the bulk LC material equal to many wavelengths, increasing the device performance sensitivity to non-uniformities in the thick LC layer. Thick LC layers also introduces significant scattering effects, negating some of the advantages of metasurface devices including aberration free focusing. By removing the bulk NLC cell, the metalens retains its 2D configuration and the phase and amplitude control occurs solely within the metasurface as opposed to the superstrate.[31] This allows for a more precise degree of control over the wavefront of light, without undue bulk phase accumulation.

Spherical gold nanoparticles (AuNPs) are an ideal candidate for approaches requiring spatially-confined heating and thermally-triggered processes.[32–35] The AuNP diameter can be selected to tune the plasmonic band into the visible region, maximizing the absorption cross-section and resultant photothermal heating. [36] Plasmonic photothermal (PPT) heating is a direct consequence of localized surface plasmon resonances (LSPR), a distinct prop- erty of metallic nanoparticles. LSPR is an electromagnetic resonance generated by light interacting with conductive nanoparticles smaller than the incident wavelength. [37,38] When these nanoparticles are embedded in thermoresponsive materials, they can be used to trigger thermally induced phenomena. [39]

In particular, AuNPs can be used to induce thermoplasmonic effects in NLCs since the behavior of thermotropic NLCs is dependent on temperature: a phase transition can be in- duced by heating (from nematic to isotropic phase) or by cooling (from isotropic to nematic phase). During the nematic-isotropic (N-I) phase transition, the liquid crystal is converted from an ordered spatial distribution where the molecules align their long axis parallel to the substrate (planar or homogeneous alignment) to a disordered (isotropic) configuration without an alignment.[25] It should be noted that the pillar structure of the metalens can disrupt the planar alignment since the LC molecules will interact with the interface at the pillar surface, leading to local deviations from the global alignment. This first-order phase transition modifies the physical properties of the liquid crystal, including the refractive in- dex, density, and wetting properties.[40] For metasurface devices embedded with NLC, the phase transition can be harnessed to enable dynamic reconfigurability, greatly expandingthe application space for these devices.

Here, we present a photothermally controlled varifocal metalens infiltrated with a mixture of 4-hexyl-4'-biphenylcarbonitrile (6CB) NLC and AuNPs, that focuses light in the visible. Dynamic tunability of the focal distance is achieved through photothermal control of the NLC phase using free-space illumination, exploiting the photo-induced phase transition due to the excitation of an LSPR mode on the AuNPs. This all-optical method offers a suitable alternative for various applications requiring compact and tunable focusing devices, avoiding mechanical actuators or dedicated electrical connections.[29] The system performance has been characterized in terms of the quality of the focus and focal length tunability range, and the reversibility and repeatability of the effect are demonstrated. The experimental results are further compared with corroborating numerical simulations of the focusing properties of the infiltrated metalens system.

## Results / Discussion

The metasurface considered here is an all-glass metalens with 1 cm diameter, focal length of 5 cm (NA = 0.1), designed to operate in air. The metalens has been previously fabricated and characterized.[14] The structure is composed of an array of fused-silica nanopillars with diameters ranging from 250 nm to 600 nm and a height of 2 $\mu$m, arranged in concentric rings, with a fixed pillar edge-to-edge distance of 250 nm. A morphological characterization of the "empty" metalens (air among the nanopillars) is reported in **Figure 1a,b**. An optical microscopy image of the central region of the metalens (**Figure 1a**) acquired between parallel polarizers (indicated with parallel arrows), which enhances the visual contrast as a function of pillar density, shows the concentric rings of densely spaced pillars interspersed with empty concentric rings, which were incorporated in the original design methodology. The structure and arrangement of the silica nanopillars within a central ring is shown via scanning electron microscopy (**Figure 1b**). A schematic of the optical setup used to characterize the focal spot and the intensity distribution of the metalens is reported in the Supporting Information (**Figure S1**). The full spatial intensity distribution has been measured within $\pm 250$ $\mu m$ from the focal length (**Figure 1c**). The focusing properties of the metalens are further characterized by the experimental point-spread function (PSF) at the focal plane (**Figure 1d**). The focusing profile, characterized by a Strehl ratio (SR) of (0.835 ± 0.002) is shown in **Figure 1e**, exhibiting diffraction limited performance. The Strehl ratio was calculated using curve fitting to the experimental PSF.

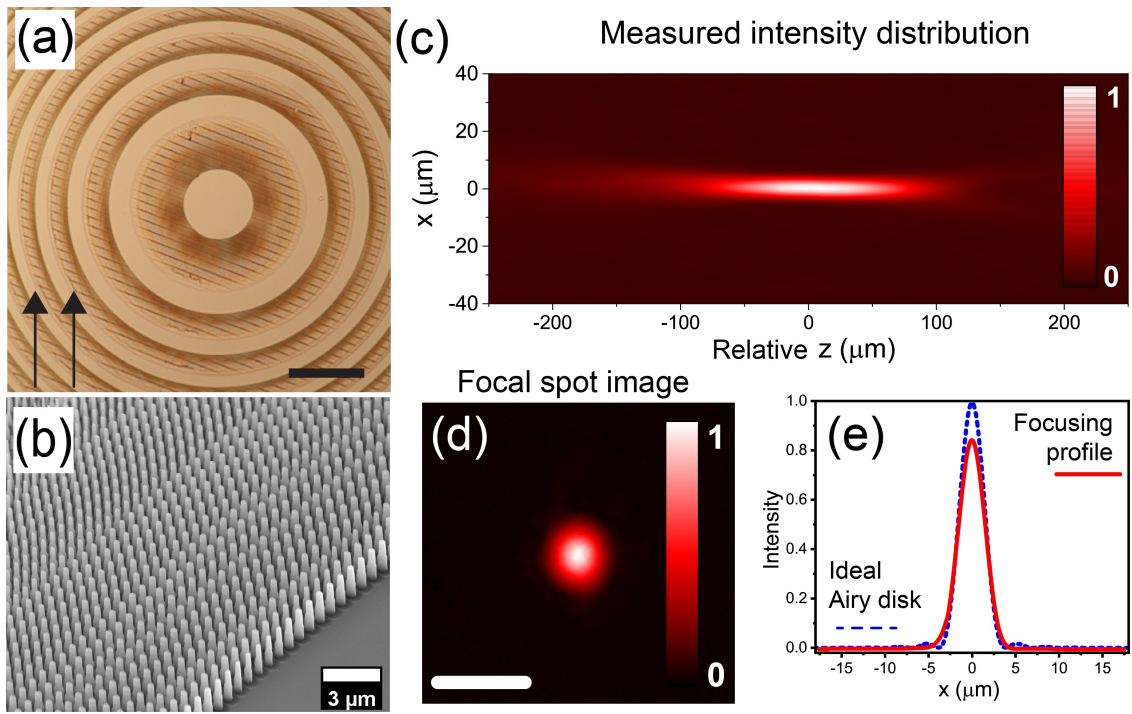

Figure 1: (**a**) Polarized optical microscopy image (with parallel polarizers (0°), shown as parallel arrows) of the central region of the empty metalens, scale bar: 200 $\mu m$; (**b**) SEM image of the pillar structure along the metalens outer edge; (**c**) the experimentally measured intensity distribution in the x-z plane about the focal point (linear intensity scale); (**d**) experimentally measured focal spot produced by the empty metalens in the focal plane, (scale bar: 10 $\mu m$) and (**e**) measured point spread function for the empty metalens, dashed lines represent the focal spot profile for ideal Airy disk. The Strehl ratio is 0.835.

The metalens was first infiltrated (**Figure 2**) with a mixture of 4-hexyl-4'-biphenylcarbonitrile (6CB) NLC and AuNPs (< 1 wt%). 6CB is characterized by a crystalline-nematic transition temperature $T_{CrN}$ = 13.64 °C and a N-I transition temperature $T_{NI}$ = 28.07 °C.[41,42] At room temperature (T∼20°C ) the birefringence of 6CB is ∼ 0.16, with ordinary and extraordinary refractive indicies of 1.53 and 1.70, respectively.[43] The temperature-dependent refractive indices of 6CB at $\lambda$=589 nm are reported in the Supporting Information (**Figure S2**). A colloidal solution of spherical AuNPs (diameter ∼ 34 nm) dispersed in ethanol was synthesized, placing the LSPR mode in the green region of the visible spectrum (see Supporting Information). The optical absorption (**Figure S3**) and photothermal conversion have been investigated. This series of photo-thermal measurements (see *Supporting Information* - **Figure S4**) revealed a strong linear correlation between laser intensity and photo-induced temperature increase of AuNPs along with optimal thermal stability.

The infiltration of the metalens with pure LC has been previously reported by Lininger et al.[44] For the current experiment, several droplets of the NLC/AuNPs mixture in the isotropic phase (heated to T > 30°C) were released around the outer edge of the metalens and left to infiltrate on a hot plate for 1 hour. During the infiltration, the NLC tends to move radially inward from the point of introduction at the outer edge of the lens after fully infiltrating the external pillared rings, and remains spatially non-uniform even after the NLC front reaches the center of the lens. The equilibrium height profile of the infiltrated NLC depends strongly on the radially varying width of the circumferential rings without nanopillars (channels), due to the wicking forces and the downward force from the droplet at the outer edge of the lens.[45,46] The infiltration of the mixture inside the metalens can be followed through an optical microscope analysis between crossed polarizers. After the 6CB+AuNPs infiltration, the metalens is visibly brighter under cross polarization due to the polarization conversion performed by the birefringent NLC (**Figure 2**). It is possible to recognize different optical textures along the edge (**Figure 2a**), inner (**Figure 2b**) and central region (**Figure 2c**) of the system. In the regions without the nanopillars a schlieren texture is observed, characterized by a series of point and line disclinations and appearing as dark brushes against a bright background. This texture is common for parallel LC alignment in un-rubbed cells.[47] It is distinct from the more uniform and darker texture observed in the regions with nanopillars.

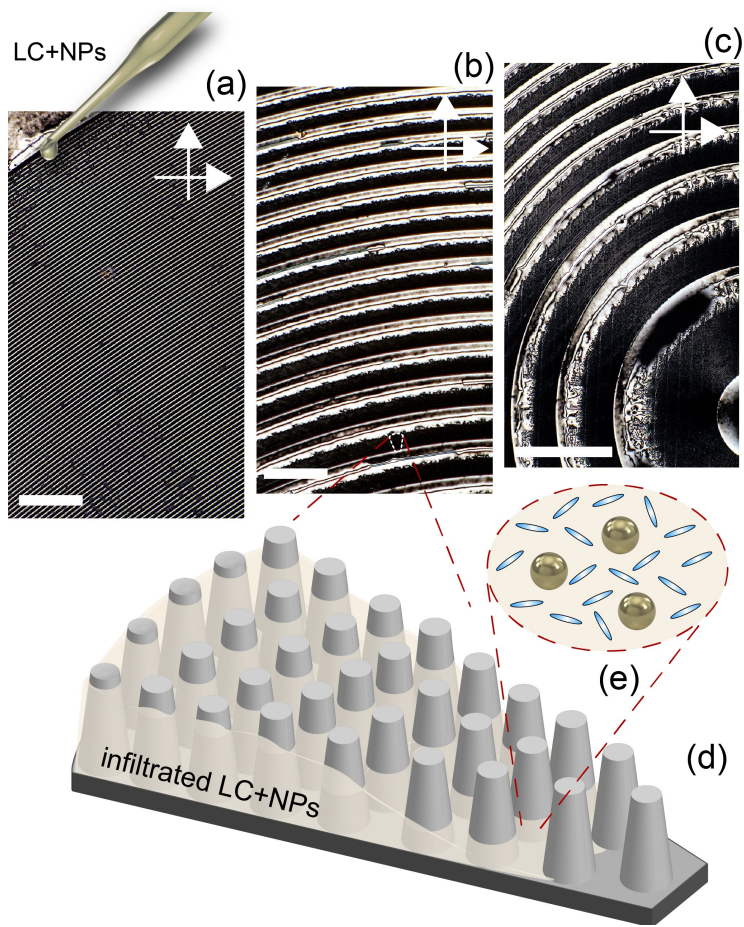

Figure 2: Polarized optical microscopy (with crossed polarizers (90°), shown as perpendicular arrows) image of 6CB+AuNPs filled metalens along the (**a**) far edge, (**b**) inner and (**c**) central region, scale bar: 200 $\mu$m. The pipette in **a** illustrates a droplet of LC placed at the edge of the lens during the infiltration procedure. (**d**) Illustration of a pillars region of the metalens infiltrated with the 6CB+AuNPs mixture.

The optical setup to control the infiltrated NLC includes a continuous wave green pump laser, on resonance with the AuNP LSPR. The light impinges at an angle of 45 degrees to the sample and the resulting temperature variations are monitored via a thermal camera, see **Figure 3a**. The green pump laser is enlarged with a beam expander (BE) to illuminate the entire surface of the metalens (beam diameter = 1 cm). A circular polarization state was used for the pump beam (including a $\lambda/4$ waveplate). This setup ensures consistent absorption for all planar NLC configurations and minimizes any strain on the NLC orientation.[48] This is necessary, since a planar configuration was observed for the infiltrated NLC (molecules aligned in the x-y plane).[25] A 570 nm high-pass filter is placed before the imaging CCD to remove the scattered pump beam from the focusing profile.

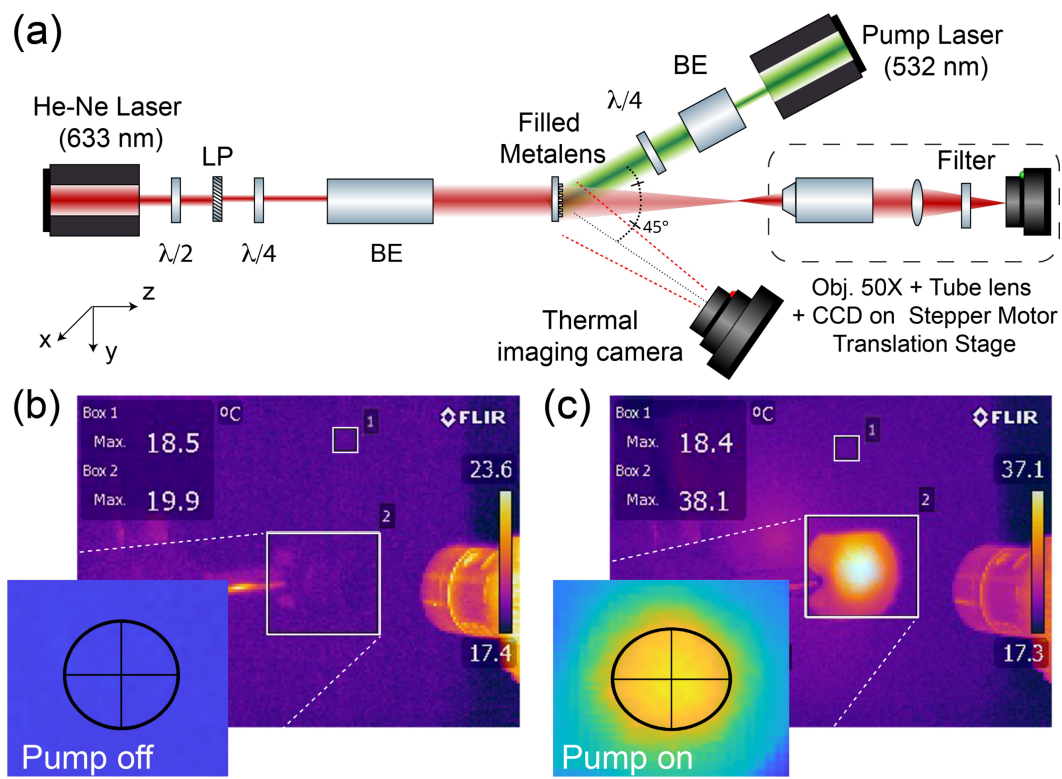

Figure 3: (**a**) Schematic diagram of the all-optical pump probe and thermographic setup: Linear Polarizer (LP), $\lambda/2$ and $\lambda/4$ waveplates, Beam Expander (BE); thermographic images of the 6CB+AuNPs filled metalens (**b**) when the pump laser beam is stopped (pump off) and (**c**) when the pump laser beam impinges on the sample (pump on, steady-state) with intensity $I$= 2.3 W/cm². Inset: temperature maps of the metalens sample (black circle) in each pump configuration, linear scale.

**Figure 3b** shows a thermographic image of the 6CB+AuNPs filled metalens (with the mounting device) when the pump beam is stopped (pump off). The two white square regions of interest, indicated as (1) and (2) in the figure, represent the two measurement regions centered on the ambient room temperature $(18.5 \pm 1.0)$°C and on the sample $(19.9 \pm 1.0)$°C, respectively. The sample is then illuminated with the pump beam (pump on) and the intensity is varied and monitored until a maximum steady-state photo-induced temperature of $38.1 \pm 1.0$ °C is reached on the sample (Box (2) in **Figure 3c**). The target temperature, obtained at a pump beam intensity of 2.3 W/cm$^2$, is greater than the NI transition temperature of T$_{NI}$ = 28.07 °C, ensuring the thermotropic transition of the NLC to the isotropic phase.

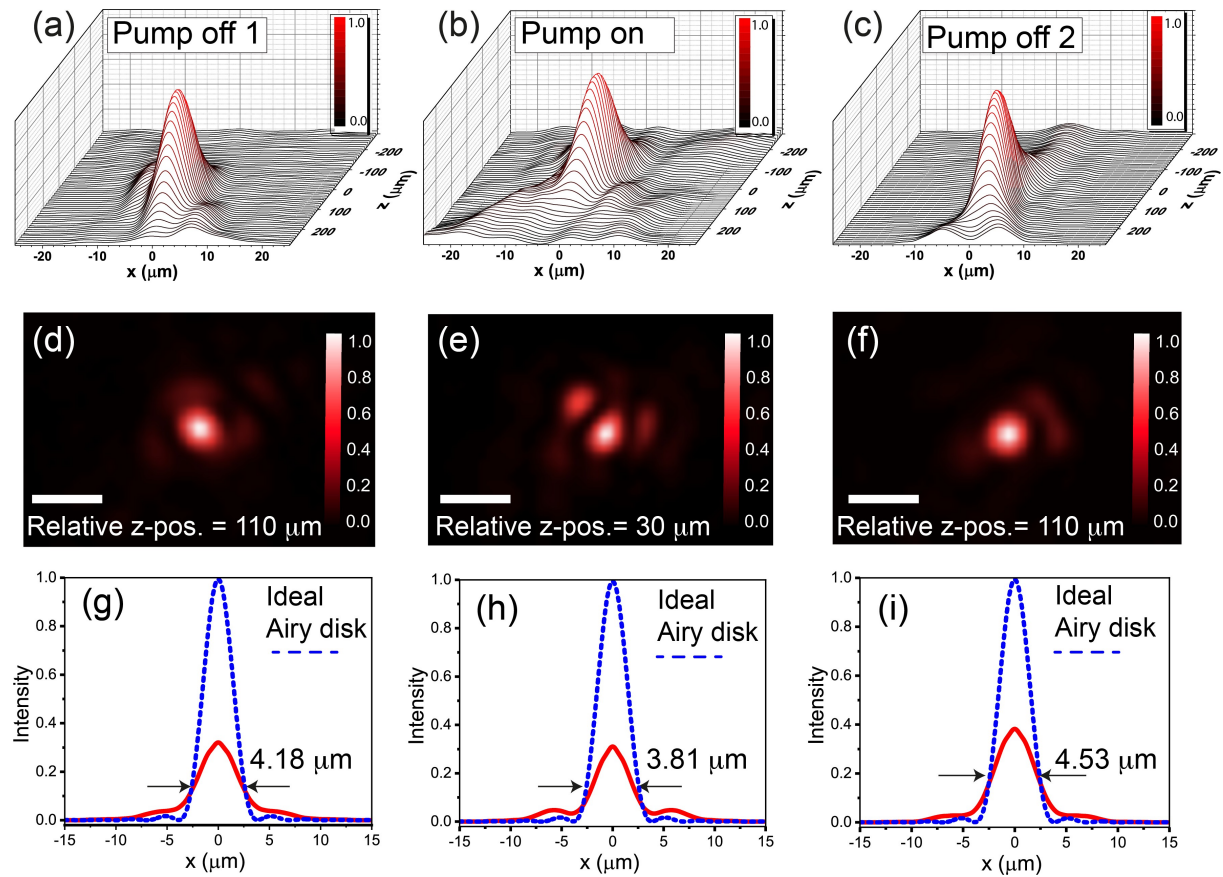

Figure 4: 3D plot of the measured intensity distributions of the 6CB+AuNPs filled metalens (**a**) before, (**b**) during and (**c**) after the exposure to the pump laser beam with (**d-f**) the corresponding intensity distributions at the focal plane (z-position is relative to the experimental zero point) for the three cases (scale bar: 10 $\mu$m), and (**g-i**) the corresponding PSFs, respectively. The blue dashed lines represent the focal spot profile for ideal Airy disk along with the experimental focusing profile in red (ideal FWHM = 3.23 $\mu$m). Experimental Strehl ratio values: (g) 0.324, (h) 0.315, and (i) 0.386.

A characterization of the focusing profile along the beam path of the 6CB+AuNPs infiltrated metalens is reported in **Figure 4**. **Figure 4a** shows the 3D reconstruction of the experimental intensity distribution of the metalens in the absence of pump beam (pump off 1), with the focal plane at the z-position of 110$\mu$m (**Figure 4d**). The z-position is given relative to the experimental zero point. The corresponding beam profile shows a FWHM of 4.2 ± 0.2 $\mu$m (**Figure 4g**). The 3D beam reconstruction is obtained by acquiring images in the x-z plane in the range −250 $\mu$m < relative $z$ < +250 $\mu$m along the optical axis, with a fixed step movement of 2 $\mu$m. Comparing the infiltrated metalens (**Figure 4d**) with the empty metalens prior to infiltration (**Figure 1c**), the infiltration is seen to decrease the quality of the focusing profile (decreasing Strehl ratio). However, the focal profile does not appear significantly altered further by the photo-thermal control procedure. The observed low Strehl ratio focusing behavior is expected since the metalens has not been specifically designed to accommodate the infiltrated NLC. Thus the infiltration disrupts the designed phase profile, and likewise the focusing performance. The quality of the focus can be further quantified by comparing the modulation transfer function for the various PSFs (presented in the *Supporting Information*).

When turning on the pump beam, a shift of 80 $\mu$m is observed in the focal distance (new *z*-position= 30$\mu$m), as well as a modification of the PSF (**Figure 4b**). Modifications in the sidelobe intensity can be observed in the focal plane 2-D intensity distribution (**Figure 4e**) while the FWHM of the central spot is decreased with respect to the pump off case. The FWHM of the focal spot is evaluated to be 3.8 ± 0.2 $\mu$m (see **Figure 4h**). The pump beam is then removed from the sample in order to probe the reversibility of the tuning effect. A 3D reconstruction of the intensity distribution, as well as the corresponding PSF and the intensity sectional profile in the x-direction at the focal plane, are reported in Figures **4c**, **4f** and **4i**, respectively. Following the heating cycle (pump off 2), the focal plane returns to the *z*-position of 110$\mu$m (**Figure 4c**), as obtained before the heating cycle (**Figure 4a**). The FWHM of the obtained PSF focal spot is 4.5 ± 0.2 $\mu$m (**Figure 4i**). The corresponding Strehl ratio values, for the three reported measurements, are SR= 0.324 ± 0.002 for pump off 1, SR= 0.315 ± 0.002 for pump on, and SR= 0.386 ± 0.002 for pump off 2. The focal length tunability for the 6CB+AuNPs infiltrated metalens system is similarly demonstrated in **Figure 5a**.

In order to further investigate the repeatability and reversibility of the tuning effect, a dynamic thermo-optical experiment was performed by monitoring the intensity in a fixed x-y plane (z-position = 30 $\mu$m) while modifying the sample temperature. The temperature was altered by controlling the exposure time to the pump beam. In the time interval 0 - 60 s a temperature of 40°C is reached, which is sufficient to induce the N-I transition (**Figure 5b** - top panel). The intensity value is obtained by integrating over the intensity distribution in the x-y plane, and the measurement is then iterated for two on/off cycles. Only one cycle is shown in **Figure 5b** - bottom panel, for clarity. Increasing the temperature of the sample from $T_1$= 20 °C to $T_7$= 40 °C, the focal spot intensity is seen to increase roughly seven-fold from the initial intensity (**Figure 5b** - bottom panel). Once the pump beam is removed, the NLC cools back to room temperature, returning to the nematic phase (blue side region of the graph). At the same time, the light intensity values are observed to return to the initial state, demonstrating the repeatability of the observed effect. A control experiment (same experimental conditions) was performed on an NLC-filled metalens without AuNPs and no significant temperature or intensity variations were observed. This result demonstrates that the photo-heat conversion is directly associated with the plasmonic photo-thermal effect of the AuNPs.

The proposed tunable metalens system can be exploited for varifocal imaging with photothermally controlled focusing. To validate the imaging ability of the tunable metalens system, a negative 1951 USAF resolution target (R3L3S1N, Thorlabs) was imaged with the metalens in place of an imaging objective (the imaging setup schematic is reported in **Figure S5**). The images (corresponding to group 4 of the resolution target) formed by the metalens are shown in **Figure 6**. Initially, when the pump laser is off, the image is in maximal focus at $z$-position = 110 $\mu$m (**Figure 6a**). When the thermo-optical control is switched on the image moves out of focus (**Figure 6b**). The image is returned to focus once the pump laser is removed (**Figure 6c**). Similarly, with the acquisition plane at the *z*-position of 30 $\mu$m, the obtained image is out of focus when the pump beam is off (**Figure 6d**), an in focus when the pump beam is on (**Figure 6e**). The image is again removed from focus when the pump beam is turned back off (pump off 2) (**Figure 6f**). The observed distance between the two well-focused image planes confirms a shift of about 80 $\mu$m in the focal length induced by the photothermal effects.

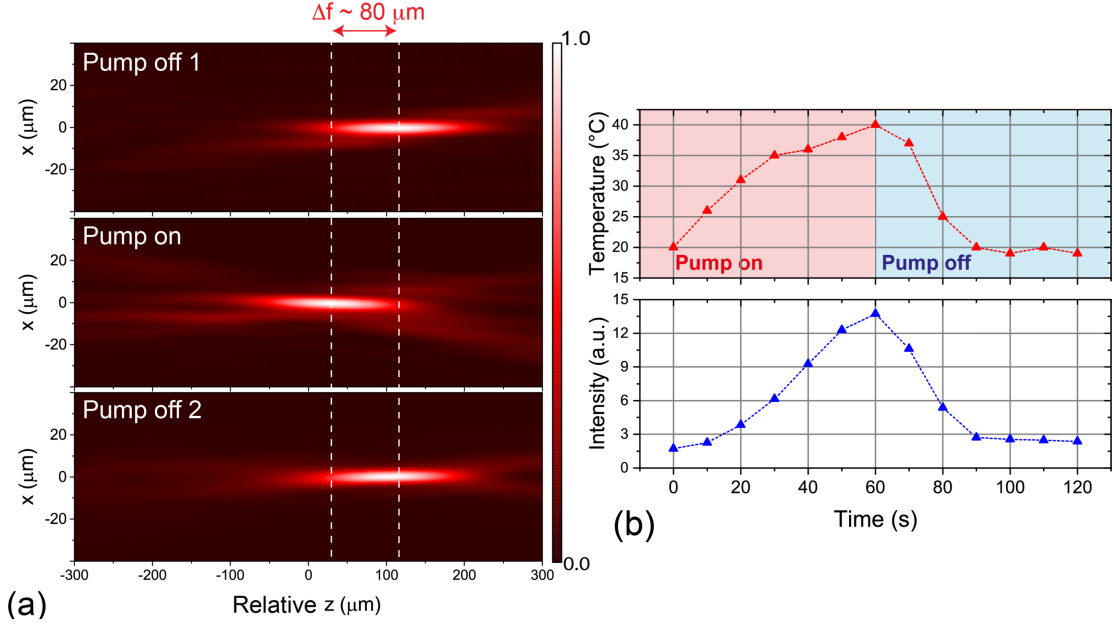

(a)

Figure 5: (**a**) Normalized measured intensity distributions along the propagation direction in the x-z plane for the 6CB+AuNPs filled metalens before (pump off 1), during (pump on) and after (pump off 2) the sample is exposed to the pump beam ($\lambda$= 532 nm). The z-position is relative to the experimental zero point. (**b**) Variations in the temperature and integrated focal spot intensity at the focal plane of the pump-active metalens (z-position = 30 $\mu$m), as the pump beam is turned on (Pump on) and then back off (Pump off) at 60 s.

## Optical Simulations

Simulations of the metalens optical properties have been conducted using a finite-difference time-domain method (MEEP). [49] Each distinct combination of pillar radius, in-plane spacing, infiltration height, and liquid crystal refractive index was simulated independently, and the amplitude and phase of the transmitted light were computed at a distance of 2$\lambda$ above the metasurface. It is seen that NLC material infiltrated into the structure greatly affects the phase transmission properties, introducing an additional phase profile superimposed upon the designed metalens hyperbolic phase profile. The individual pillar responses were configured with the known metalens geometry to obtain the total transmitted near-field, and vectorial diffraction propagation was applied to calculate the far-field intensity distribution near the focal point. The metalens was simulated in free space (no infiltration), and with 6CB infiltration at both the ordinary ($n = 1.53$) and isotropic ($n = 1.58$) refractive indices. More details on the simulation procedure can be found in the Supporting Information (*simulation procedures*).

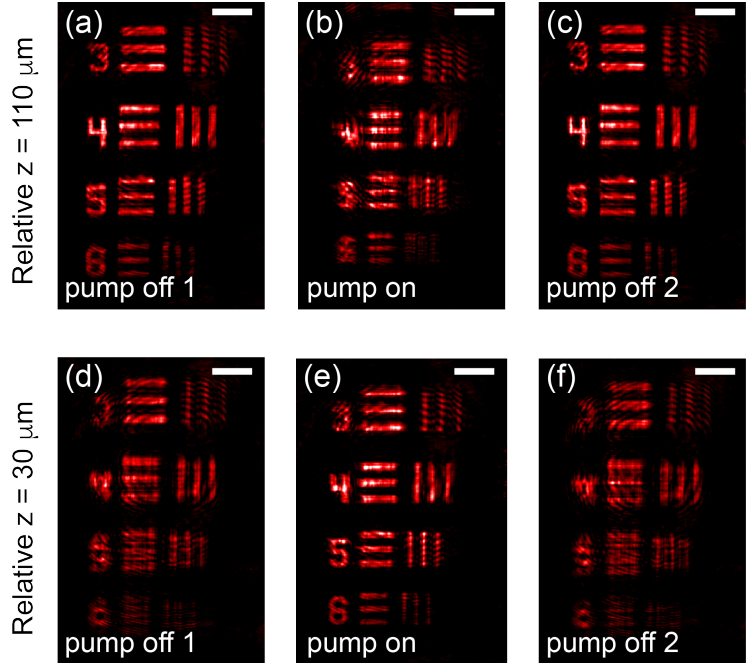

Figure 6: Experiments showing the tuning of the imaging distance of the metalens. The target image (USAF 1951 target, group 4), at relative $z$-position = 110 $\mu$m results to be (**a**) in focus when pump beam is off (pump off 1), (**b**) out of focus for pump beam on and (**c**) again in focus when pump beam is off (pump off 2). In similar manner, the target image, at relative $z$-position = 30 $\mu$m results to be (**d**) out of focus when pump beam is off, (**e**) in focus for pump beam on and (**f**) again out of focus when pump beam is off. The scale bar is 400 $\mu$m. Note that some speckle is present due to the coherent illumination.

The infiltration profile is dependent upon the hemiwicking properties of the metasurface. This process describes the spreading of liquid on a patterned surface as driven by sponta-neous capillary action.[50,51] The rate and degree of spreading depends on both the interfacial interaction energy for the liquid-glass and liquid-air interfaces, as well as the viscoelasticity of the NLC.[45] In addition, the structured nanoscale geometry of the surface critically influences the hemiwicking (infiltration) process.[46] In these simulations, we have assumed that the NLC follows an increasing radial infiltration profile, as illustrated in **Figure 7a,b**. Theprofile increases quadratically from the center of the metalens, leveling off at 80% (4 mm) of the total radial distance. This choice of filling profile is consistent with the geometry of the metasurface, considering the radially increasing spatial frequency of the pillar bands com- bined with the hemiwicking properties (**Figure 7a**) and experimental phase compensator measurements of the liquid crystal infiltration height (details in the Supporting Informa- tion).[44] We acknowledge that this choice of filling profile is not distinct, and experimental deviations from the profile are probable. However, the quadratic profile serves as a useful low-order model to parametrize the device geometry. The maximum height of the infiltrated nematic phase of 6CB is assumed to be 40% of the total pillar height ($0.80\mu m$), corresponding to a relatively low volumetric degree of infiltration.

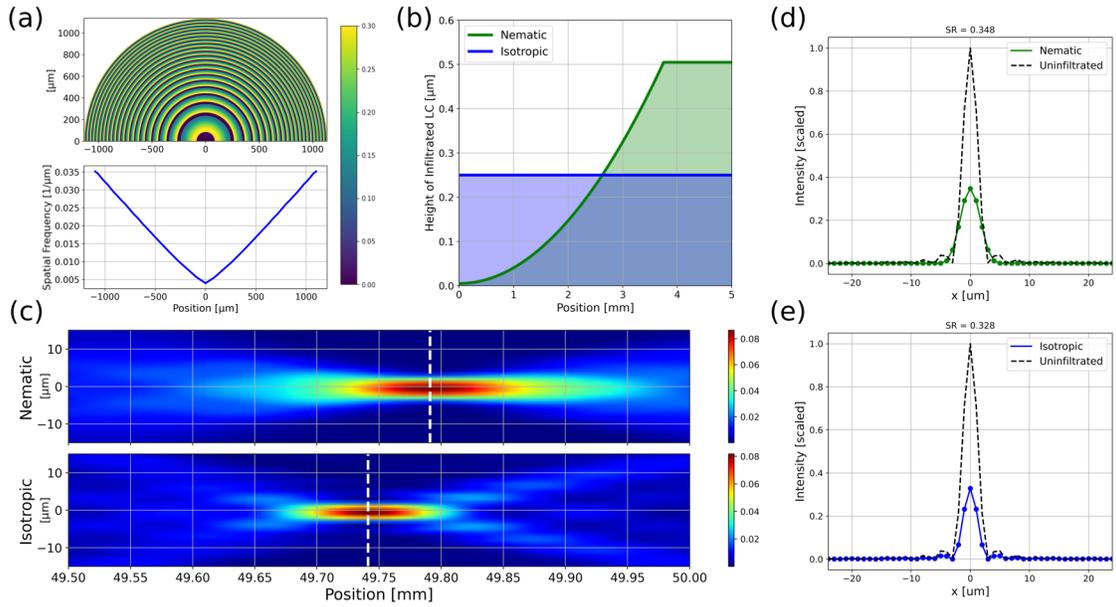

Figure 7: Optical simulations of the infiltrated metalens, including the reconfigurability associated with the N-I transition. (**a**) Illustration of the metalens structure. (Top) Color map of the radius of the fabricated pillars for the central 1 mm radial section of the metalens.Note the concentric bands of pillars and non-pillared regions. The colorbar represents the nanopillar radii in $\mu$m. (Bottom) The spatial frequency of the pillar bands increases with radial distance from the center. This corresponds to decreasing the widths of the non-pillared rings, affecting the infiltration. (**b**) The filling profiles considered in the simulations are shown for the entire metalens radius. In the nematic phase we assume the LC to follow a quadratically increasing radial infiltration profile, reaching a maximum height of 0.8 $\mu$m at 4 mm from the center. We also assume the isotropic phase LC is at a constant infiltration height (0.55 $\mu$m) with the same volume as the nematic phase LC. The theoretical infiltration profile is compared with experimental phase compensator measurements of the infiltrated LC in the nematic phase. (**c**) Comparison of the far-field spatial intensity distributions calculated for each of the infiltration profiles above. Focusing behavior is illustrated for both the nematic (top) and isotropic (bottom) simulations. The peak intensity (arb. units) is similar for both focal spots, as is the depth of focus. A clear shift of 60$\mu$m is demon- strated between the two simulations. (**d,e**) Comparison of the PSF for both nematic (**d**) and isotropic (**e**) simulations. Each case is compared with the PSF obtained from a similar simulation of the un-infiltrated metalens. A clear decrease in peak intensity with respect to the un-infiltrated focusing abilities is observed. This is reflected in the calculated Strehl ratios of 0.296 and 0.441, respectively. The theoretical PSFs are qualitatively similar to the experimental results.

The properties of the NLC are known to change when undergoing the first order nematic-isotropic (N-I) phase transition, as predicted by Landau-de Gennes theory.[52–54] Here, we assume that the effect of the N-I transition is to alter the refractive index ($n = 1.58$) and decrease the NLC viscoelasticity and surface tension at the liquid-glass interface. This can lead to a 'smoothing' of the radial infiltration profile in the isotropic phase, which we approximate as a constant height of infiltration at the same total volume (28%, $0.55\mu m$). This modification of the infiltration profile affects the additional phase profile introduced by the infiltrated NLC. Again, we acknowledge that a completely smooth infiltration is not necessarily realistic to the experimental case since the infiltrated profile depends on the specific experimental details of the infiltration. Any deviations from the constant profile will likely increase aberrations in the far field. However, we propose that the combination of coexisting phase profiles from the fabricated lens and the infiltrated NLC is one possible explanation for the observed reconfigurable focusing behavior. This point is further discussed in the Supporting Information (*Mechanism of Coexisting Phase Profiles*).

The far field intensity profiles and PSFs corresponding to these systems are shown in **Figure 7c, d-e**, respectively. Considering the N-I transition, we observe a shift in the focal point of $\sim 60\mu m$ along the optical axis when passing from the nematic to isotropic phase. This focal point shift is similar to the experimental result (0.12% compared to 0.16% of the nominal focal length, respectively, and in the same axial direction). Interestingly, the depth of focus and maximal intensity of the main focal spot remain largely unchanged throughout the transition. Non-ideal focusing behavior in the form of aberrations near the focal spot should be noted in both cases. Similar intensity aberrations are observed in the experimental data, although the specific location and intensity of these effects are highly dependent on the specifics of the infiltrated profile and unlikely to be captured by simulation. The resulting PSFs are shown in **Figure 7d-e**, along with the experimental profiles. A clear focusing behavior is observed, however the infiltration of LC into the metalens structure is seen to affect the quality of the focus. The theoretical focusing efficiencies (defined as the power passing through 3 Airy disk diameters at the focal plane, relative to the incident power) are calculated as 18% and 20%, respectively. In both cases the infiltration leads to an increase in the sidelobe intensity. This is reflected in the Strehl ratios of 0.30 and 0.44 for the nematic and isotropic profiles, respectively. These values are similar to the experimental Strehl ratios of 0.36 and 0.32 ($\pm$ 0.06 and 0.12, respectively). The modulation transfer functions for the calculated PSFs are shown in the *Supporting Information*. The consistently low Strehl ratios in both theory and experiment are reflective of non-ideal focusing behavior in the metalens system. This is expected behavior, and occurs because the metalens utilized in this work has not been specifically designed to integrate the phase induced by the NLC infiltration, causing the infiltrated NLC to disrupt the designed phase profile.[44]

# Conclusions

We report a nano-structured metalens device permeated with a nematic liquid crystal and gold nanoparticle solution. A dynamic, reversible, and all-optical control of the lens focal length is demonstrated via a light induced thermo-plasmonic effect. A total focal shift of 80 $\mu$m in the focal length is observed. This work is a significant step towards all-optical tunable metalenses by expanding the working range from the near-infrared to the visible regime. In comparison to existing systems, this device offers an all-optical tunable metalens in which the NLC+AuNPs mixture is directly infiltrated into the metasurface structure without the necessity of assembling bulky LC-cell. This leaves the 2D flat surface of the metalens unchanged, a bulk LC superstrate. Furthermore, this metalens system is fully compatible with conventional CMOS fabrication techniques and is simple to control in comparison to tunable devices based on mechanical stretching or other control methods.

The metalens device has been experimentally demonstrated for imaging. Future applications include adaptive vision, bio-imaging, display applications where varifocal or multi-planeimaging is desired. The imaging results could be improved by tuning the fabricated silica structure to better accommodate the optical properties of the infiltrated LC, increasing the quality of the focal spot under various illumination conditions. In the current system, the phase profile is offset from the designed target by the infiltrated LC material. In an ideal system the pillar structure would be designed such that the target phase profile is achieved after the LC is infiltrated with a typical height profile. To improve the focal point shift distance, pillar geometries can be designed which maximize difference in transmitted phase due to the LC phase transition. Altering the pillar arrangement will also alter the hemiwicking properties of the nanostructure, which could be designed for specific infiltration profiles. Optimization methods, in particular machine learning approaches, would be useful for this task.

Finally, the experimental results are compared with corroborating numerical simulations of the focusing properties of the infiltrated metalens device. These simulations offer an explanation of the reconfigurable focusing properties of the device, however some assumptions of the model (particularly the specific infiltration profile) are not distinctive, and do not necessarily reproduce the experimental case.

# Methods/Experimental

## Liquid Crystal and Nanoparticle Preparation.

Citrate-stabilized AuNPs with an average diameter of $34 \pm 2$ nm were synthesized following the well-developed kinetically controlled seeded growth method,[55] via the reduction of $HAuCl_4$ by sodium citrate. All chemicals were purchased from Sigma-Aldrich. The synthe- sized AuNPs were purified by ultrafiltration (Vivaspin20 equipped with a 100 kDa membrane) and then dispersed in spectrofluorimetric grade ethanol with a final concentration of 2 nM. The AuNPs solution was added to 4-hexyl-4'-biphenylcarbonitrile (6CB) NLC at 1 wt %, and the solved evaporated to produce the NLC + AuNP mixture. More details on the AuNP and NLC preparation can be found in the *Supporting Information*.

## Experimental Heating and Focusing Measurement.

Heating of the infiltrated lens was performed with a continuous wave green laser ($\lambda = 532$ nm, Verdi, Coherent) with circular polarization, at $I = 2.3$ W/cm$^2$. The light impinges upon the sample at 45 degrees angle of incidence. The temperature variations were monitored with a thermal imaging camera (E40 *FLIR*). The focusing was measured for 633 nm incident light (He-Ne, JDS Uniphase 1137P). A repositionable microscope setup with a 50x objective (Leica 566040, N.A.= 0.5) was utilized to measure the far-field intensity pattern.

## Optical Simulations.

Simulations of the metalens optical properties were conducted using MEEP, an open source finite-difference time-domain (FDTD) method package. [49] All combinations of the pillar radius, in-plane spacing, infiltration height, and liquid crystal refractive index were simulated independently. The amplitude and phase of the transmitted light were computed at a distance of $2\lambda$ above the metasurface. The far-field focusing pattern was simulated by combiningthe individual pillar light transmission for the known pillar geometry and infiltration profile, and applying a far-field transformation. More details on the simulations can be found in the *Supporting Information*.

# Acknowledgement

We acknowledge support from the Ohio Third Frontier Project "Research Cluster on Surfaces in Advanced Materials" (RC-SAM) at Case Western Reserve University. A.L. and G.S. acknowledge financial support from the NSF Grant no. 1904592 "Instrument Development: Multiplex Sensory Interfaces Between Photonic Nanostructures and Thin Film Ionic Liquids". G.P. acknowledges financial support from the "AIM: Attraction and International Mobility" - PON R&I 2014-2020 Calabria. G.N. acknowledges financial support from the "Dottorati innovativi a caratterizzazione industriale" - PON R&I FSE-FESR 2014-2020. A.G. acknowledges financial support from the "NLHT - Nanoscience Laboratory for Human Technologies" - (POR Calabria FESR-FSE 14/20). S.W.D.L. is supported by A*STAR Singapore through the National Science Scholarship Scheme. The Harvard University authors were partially supported by the AFOSR MURI Grant # FA9550-21-1-0312. This work was performed in part at the Cornell NanoScale Science Technology Facility, a member of the National

Nanotechnology Coordinated Infrastructure (NNCI), which is supported by the NSF (Grant NNCI-1542081). This work was also performed in part at the Center for Nanoscale Systems (CNS), a member of the NNCI, which is supported under NSF Award 1541959. CNS is part

## Supporting Information Available

Supporting Information. Schematic diagram of the optical setup used to characterize the focal spot and the intensity distribution of the metalens. Temperature-dependent refractive indices of 6CB at $\lambda$=589 nm. Synthesis and characterization of the AuNPs. Photother- mal characterization of the AuNPs. Schematic diagram of the full experimental apparatus. Experimental and theoretical MTF comparison. Simulation procedures. Mechanism of co-existing phase profiles. Manipulating the focal point tunability by varying the infiltrate refractive index. Experimental measurement of liquid crystal height. Comparison with some other similar tunable metalens systems.